\documentclass{iau} 
\usepackage{graphicx}

\title[Radio Supernova 1987A] %% give here short title %% 
{Radio Observations of Supernova 1987A}

\author[L. Staveley-Smith et al.]   
{L. Staveley-Smith$^{1,2}$, T.M. Potter$^{1}$,
G. Zanardo$^1$,\\ 
B.M. Gaensler$^{2,3}$ \and C.-Y. Ng$^{4}$}

\affiliation{$^1$International Centre for Radio Astronomy Research,
University of Western Australia, Crawley, WA 6009, Australia \\ 
email: {\tt Lister.Staveley-Smith@icrar.org}\\[\affilskip] 
$^2$ARC Centre of Excellence for All-sky Astrophysics (CAASTRO)\\[\affilskip]
$^3$Sydney Institute of Astronomy, School of Physics, The University of Sydney, NSW 2006, Australia \\[\affilskip]
$^4$Department of Physics, The University of Hong Kong, Pokfulam Road, Hong Kong}

\pubyear{2013} \volume{296}  %% insert here IAU Symposium No.
\pagerange{119--126}
 \jname{Supernova Environmental Impacts}
\editors{A. K. Ray \& R. McCray, eds.} 

\begin{document}

\maketitle

\begin{abstract} 
Supernovae and their remnants are believed to be
prodigious sources of Galactic cosmic rays and interstellar dust.
Understanding the mechanisms behind their surprisingly high production
rate is helped by the study of nearby young supernova remnants. There
has been none better in modern times than SN1987A, for which radio
observations have been made for over a quarter of a century. We review
extensive observations made with the Australia Telescope Compact Array (ATCA) at centimetre
wavelengths.
Emission at frequencies from 1 to 100 GHz is dominated by synchrotron radiation from
an outer shock front which has been growing exponentially in strength from day 3000, and
is currently sweeping around the circumstellar ring at about 4000 km s$^{-1}$.
Three dimensional models of the propagation of the shock into
the circumstellar medium are able to reproduce the main observational features
of the remnant, and their evolution. We find that up to 4\% of the electrons encountered 
by the shock are accelerated to relativistic energies. 
High-frequency ALMA observations will 
break new ground in the understanding of dust and molecule production.
%% add here a maximum of 10 keywords, to be taken form the file
%% <Keywords.txt>
\end{abstract}

\firstsection % if your document starts with a section,

\section{Overview}

As the brightest supernova since Kepler's of 1604, SN1987A has been a Rosetta Stone in our
understanding of the physics of Type II supernova explosions. It was the first supernova with a known 
progenitor, and the brightest since the invention of the telescope. It has therefore allowed  
a number of unprecedented studies including: detailed comparison with models of the very 
final stages of stellar evolution; the detection of the first neutrinos from an extrasolar
source, providing evidence for the formation of a neutron star (\cite{Vissani10}); 
measurement of absolute distance using 
the light travel time method (\cite{Panagia91}); the probing and analysis of the circumstellar medium excited by the ultraviolet flash; measurement of the radioactive decay products of the explosion; study of the 
interaction of the expanding blast wave with the anisotropic circumstellar medium
laid out by the stellar winds from various stages of the progenitor star; and formation of dust 
and molecules in the cooling ejecta. The fact that, to date, over 2000 refereed papers have 
been written about SN1987A, or have mentioned SN1987A in their abstracts, is testament to
the significance of the event.

In this paper, we discuss radio observations in the two and a half decades following the explosion. These observations
are particularly sensitive to detailed shock physics, magnetic field amplification and circumstellar structure, and have allowed
us to clearly follow the evolution of a Type II supernova into the important supernova remnant (SNR) phase.

\begin{figure}[t]
% \vspace*{-2.0 cm}
\begin{center} \includegraphics[width=\textwidth]{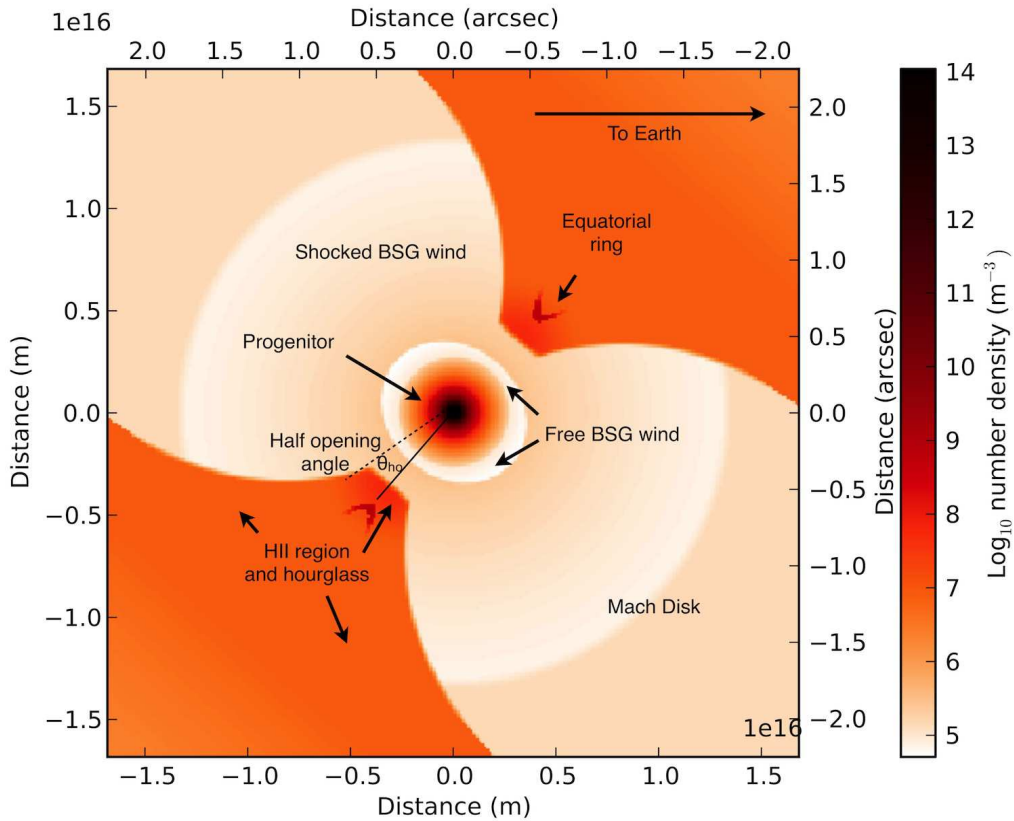} 
 \vspace*{1.0 mm}
 \caption{A model of the initial `hourglass' environment prior to explosion. The environment
 was created by the progenitor star in its evolution from red supergiant to blue supergiant (BSG).
 The supernova shock is currently at the position of the equatorial ring the ring is 
 perpendicular to the plane of the paper (\cite{Potter12}). A Mach disk is formed at
 the point where the BSG wind again goes supersonic. The axes are labelled in arcsec and in units of
 $10^{16}$m.} \label{lssmodel} \end{center} \end{figure}

\vspace*{-1.0 mm}
\section{Environment}

The initial interaction of the SN1987A blast wave 
with the circumstellar medium resulted in the generation of a burst of synchrotron radiation 
lasting a few days
(\cite{Turtle87}; \cite{Storey87}). Radio (and X-ray) observations of other, more distant, type II supernovae 
have been used to compare progenitor mass-loss rates using the theoretical framework laid out 
by \cite{Chevalier82}, and to probe the uniformity of the mass loss. Similar studies on SN1987A were
consistent with the observation that the progenitor was a blue supergiant, with a fast but tenuous 
stellar wind at the time of explosion.
However SN1987A was notable in that the luminosity of this initial radio burst, which peaked at about day 3, was around $10^4$ times weaker
than powerful type II supernovae such as SN1993J. Nevertheless, following the discovery by
the ESO New Technology Telescope (\cite{Wampler90}) and the HST of a ring-like structure around SN1987A,
there was some expectation of a re-brightening of the radio emission. This was seen at day 1200
by the MOST and ATCA telescopes (\cite{Staveley92}). 

As a result of early optical and radio observations, and studies of the light echoes of the 
explosion itself (\cite{Crotts89}), a detailed picture of the environment around the supernova has been constructed. The main features of the circumstellar environment are illustrated in Fig.\ref{lssmodel}. The well-known equatorial ring, which defines the waist of an hourglass feature, is perpendicular to the plane of the paper. The 
northernmost edge of the ring is closest to Earth. Within the hourglass, there are (pre-supernova) regions
which signify both freely expanding wind and shocked wind. It is at the interface of the two where it is 
believed that the first radio emission was generated (\cite{Chevalier95}).

\begin{figure}[t]
% \vspace*{-2.0 cm}
\begin{center} \includegraphics[width=0.82\textwidth]{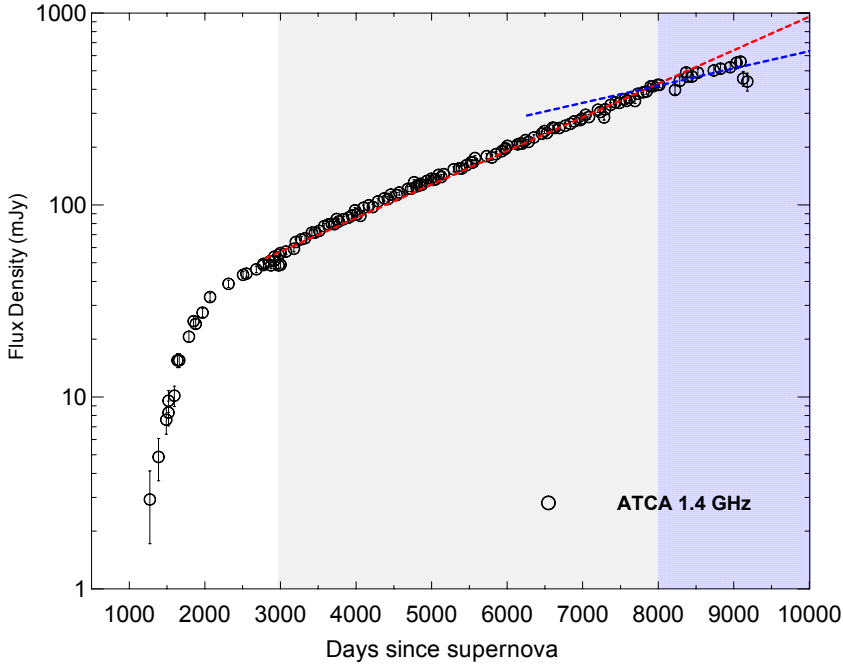} 
% \vspace*{-1.0 cm}
 \caption{Radio observations at 1.4 GHz have been conducted with ATCA at intervals of 4-6 weeks
 since the remnant was re-detected around day 1200. Its evolution is divided into three phases: (a) a linear phase up to day 3000; (b) an exponential phase to day 8000 (light shading); and (c) a slower exponential phase from day 8000 (darker blue shading). 
 The red dotted line is the fit of \cite{Zanardo10}.} 
 \label{lssflux} \end{center} \end{figure}

\section{Evolution}

After re-emergence of the radio remnant at day 1200, the flux density was seen to increase
linearly with time until day 4500, when small departures were evident (\cite{Manchester02}).
In datasets from day 3000 to 8000, a better description is an exponential increase
(\cite{Zanardo10}), perhaps reflecting the exponential nature of the particle acceleration 
process and/or the increasing area of the expanding shock front. Observational data from the ATCA at 1.4 GHz is shown 
in Fig.\ref{lssflux}.

\begin{figure}[t]
%\vspace*{1.0 mm}
\begin{center} \includegraphics[trim=2.0mm 2.0mm 0.0mm 0mm,width=0.95\textwidth]{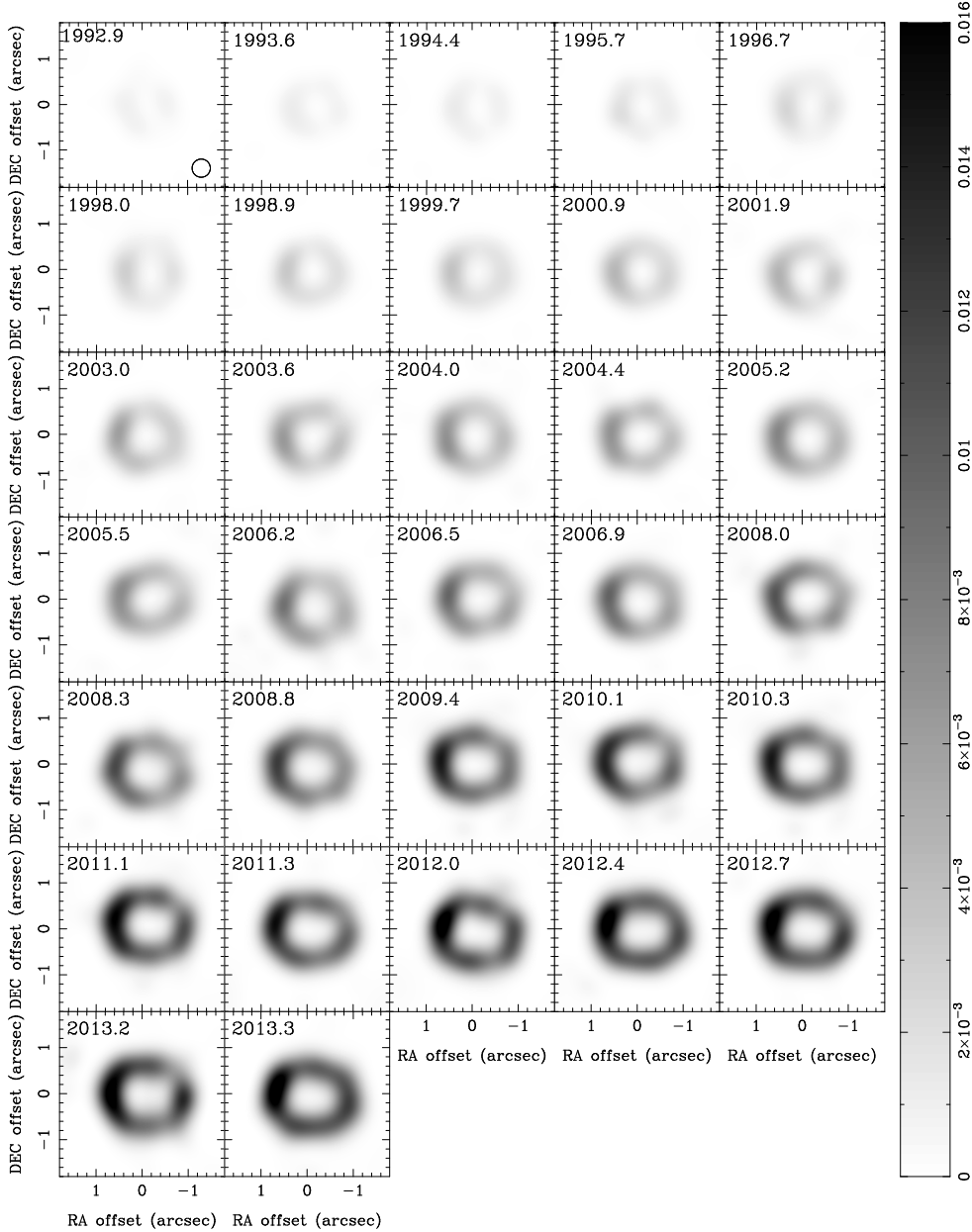} 
 \vspace*{1.0 mm}
 \caption{Radio observations at 9 GHz have been conducted with ATCA at half-yearly or yearly intervals
 since the SN1987A remnant was first resolved in 1992. The above panels illustrate: (a) the
 strengthening radio emission; (b) the change from a circular (torus) to an elliptical (equatorial
 ring) morphology; and (c) the increasing diameter of the remnant. The year (CE) of the observation
 is listed in the top-left corner of each panel. Note the asymmetric
 nature of the remnant. Data from \cite{Ng13}.} \label{lsspanel} \end{center} \end{figure}
 
Similar exponential evolution was seen in the X-rays, though the detailed temporal behaviour
has been quite different (\cite{Helder13}). This is now understood to be due to the different emission regions
and the dominance of X-ray thermal radiation over synchrotron radiation dominant at radio wavelengths.
Interpretation of the radio light curve has been assisted by high-resolution radio observations with
the Australia Telescope Compact Array (ATCA), particularly when the remnant became strong enough
to apply the technique of `super-resolution'. This allows a resolution increase by a factor of 1.5 to 2
for a compact source like SN1987A with the available signal-to-noise ratio. 
A compilation of all 9 GHz full uv-coverage ATCA images is shown in Fig.\ref{lsspanel}. Firstly, the
rapid brightening of the whole remnant is apparent. This reflects the radio light curve shown
in Fig.\ref{lssflux}, although at 9 GHz the increase is slightly steeper, reflecting the
flattening of the radio spectrum over time. Secondly, on close inspection, the growth in the size of the remnant can be
measured, reflecting the current expansion rate of 4000 km s$^{-1}$ (\cite{Ng08}). 
Notably, the expansion rate was 
much higher prior to switch-on of the radio emission at day 1200. This difference is explained as
an inner cavity in models of the pre-supernova medium as shown in Fig.\ref{lssmodel}. 
Finally, it is apparent that the remnant is one-sided, with the eastern half being consistently $\sim30$\% brighter until about 2007, after which the asymmetry decreases. The most likely explanation
of this is explosion asymmetry. \\

 \vspace*{-5 mm}
\section{Models}

A number of hydrodynamic models have explored the propagation of the SN1987A shock front through
the circumstellar medium. These have been used to predict dates for the collision of the shock front with the ring
(\cite{Borkowski97}) and have also been used to explain the radio evolution of the remnant
(\cite{Berezhko06}). However, most simulations
have been one-dimensional and do not capture the inherent asymmetry of the pre-supernova environment.
We have therefore attempted to model the hydrodynamical propagation of the shock front in three dimensions.
We do this using an initial model which was motivated by observations and itself created from a 
plausible hydrodynamic evolution for a combined red/blue supergiant system. The initial model is
that of Fig.\ref{lssmodel}. Further details (densities, temperatures, dimensions) are given in \cite{Potter12}. 

\begin{figure}[t]
% \vspace*{-2.0 cm}
\begin{center} \includegraphics[width=8.0cm]{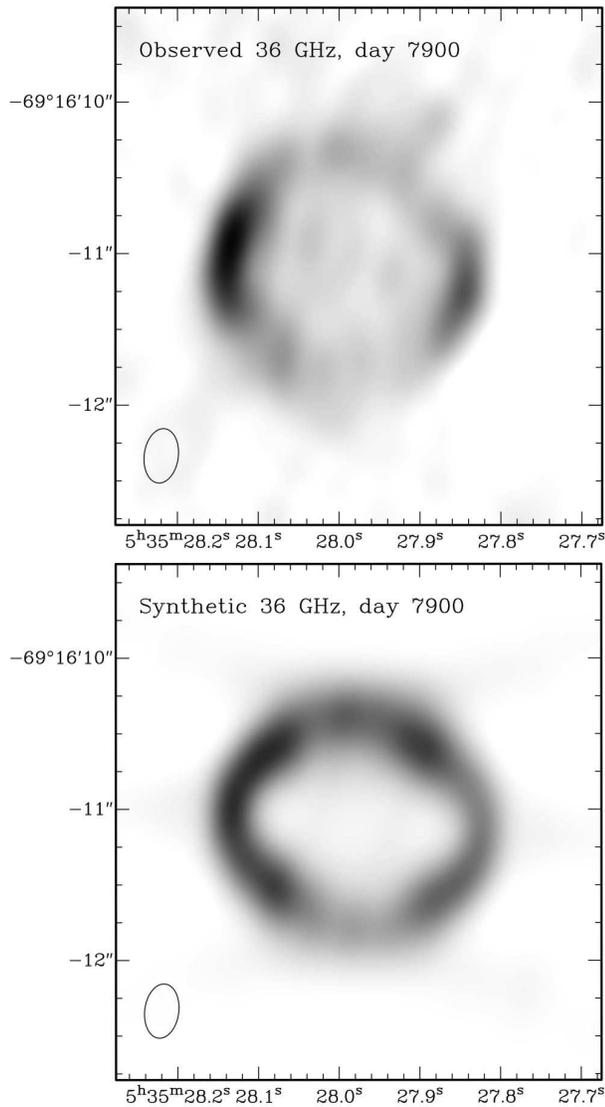} 
% \vspace*{-1.0 cm}
 \caption{{\it Top:} A 36-GHz ATCA image of the radio remnant of SN1987A on day 7900 (\cite{Potter09}).
{\it Bottom:} multi-dimensional hydrodynamic DSA model of the remnant at the same frequency and on the
same day (\cite{Potter12}).} \label{lssimage} \end{center} \end{figure}

The radio emission is calculated
from the hydrodynamic simulation using sub-grid physics, although it was not possible to include any back-reaction
from the cosmic rays onto the shock front. Thermal electron are injected into the shock and 
assumed to be accelerated to high energies by the process of Diffusive Shock Acceleration (DSA). The
electron energy spectrum and the magnetic field energy are used to calculate the synchrotron
emissivity, which adiabatically decays after advection from the shock. As the simulations are in three dimensions, we can create synthetic radio images at any time and frequency. An example is the 
synthetic 36 GHz image at day 7900, which is plotted alongside an actual ATCA image from \cite{Potter09} at the same frequency in Fig.\ref{lssimage}. Parameters such as diameter, asymmetry and axis ratio
can be calculated, and seem to accurately reproduce the past evolution of the remnant. Future 
predictions (Potter et al. in preparation) include a reversal of the asymmetry of the remnant.
However, due to the highly non-linear nature of the DSA mechanism, it is more challenging to reproduce the exact shape of the radio light curve of Fig.\ref{lssflux}. Nonetheless, our best fit suggests that the injection rate of those thermal electrons which are accelerated to cosmic
ray energies must be around 4\%. This is remarkably efficient and gives an insight into the
importance of SNRs as sources of highly energetic particles in galaxies.

\section{Future}

SNRs are believed to be important sources of cosmic rays in galaxies. These particles are
important in facilitating the conversion of atomic gas into molecular form by providing charge to dust 
grains deeply embedded in these clouds. They also provide interstellar pressure that slows the accretion of cool gas onto galaxies.
Cosmic ray acceleration processes can be further studied by tracking the evolution of the radio emission
at the forward shock during the ongoing transition of SN1987A into an SNR. 

High energy particles are also created by
magnetised spinning neutron stars. Although no pulsar has yet been detected, it is believed
that a neutron star was created during the initial explosion. Evidence for this comes from the $\sim21$
non-background neutrinos seen by the Kamiokande II, IMB and Baksan detectors about 2 hrs before the visible flash (\cite{Vissani10}).
\cite{Zanardo13} suggest that the conditions may now be favourable for detection of a pulsar.
Indeed, tentative signs of a flat spectrum central component have been found in 
previous multi-frequency radio studies (\cite{Potter09}; \cite{Zanardo13}).

Finally, sub-mm observations indicate the presence of large amounts of cold dust in the ejecta of
SN1987A (\cite{Matsuura10}). This is exciting as it implies that dust formation in 
supernovae is
probably more important than previously believed, and perhaps resolves the puzzle of the rapid appearance of large 
quantities of dust at high redshift. Although not the subject of this paper, multi-frequency
and spectral-line observations with ALMA will be crucial in the measurement of dust and gas mass and temperature,
gas kinematics and in-situ dust and molecule formation.

\section*{Acknowledgements}
Parts of this research were conducted by the Australian Research Council Centre of Excellence for All-sky Astrophysics (CAASTRO), through project number CE110001020.

\end{document}